\documentclass[prd,aps,showpacs,preprintnumbers,amssymb]{revtex4}

\usepackage{color}
\usepackage{epsf}

\def\e3p{$\eta \rightarrow 3 \pi$}

\begin{document}
\title{%
\hfill{\normalsize\vbox{%
\hbox{}
 }}\\
{About the electroweak vacuum and its connection with the quark condensates}}

\author{Renata Jora
$^{\it \bf a}$~\footnote[2]{Email:
 rjora@theory.nipne.ro}}

\affiliation{$^{\bf \it a}$ National Institute of Physics and Nuclear Engineering PO Box MG-6, Bucharest-Magurele, Romania}

\date{\today}

\begin{abstract}
We make a change of variable in the standard model Higgs field by a fermion operator and show that the latter is responsible for the electroweak vacuum. By computing the vacuum expectation value of this fermion operator in the path integral formalism we determine a relation among the quark vacuum condensates, the quark masses and the intrinsic scale of the theory. We show that the heavy quark vacuum condensates do not justify the  hypothesis of dynamical electroweak symmetry breaking with only standard model fermions.

\end{abstract}
\maketitle

\section{Introduction}

After decades of experimental searches the standard model Higgs boson was finally discovered by the LHC experiments \cite{Atlas}, \cite{CMS}  and its mass was established at $m_h=125.09$ GeV. This great experimental success was however not accompanied by any hint of beyond of the standard model physics. With each passing year the robustness and precision of the standard model is confirmed and the parameter space of the models that extend the existing theory is pushed farther towards the ${\rm TeV }$ scale. In essence there are three main alternatives to the standard electroweak symmetry breaking:  dynamical electroweak symmetry breaking \cite{Weinberg1}-\cite{Chivukula}; supersymmetry \cite{Nilles}-\cite{Kane} and extra dimensions \cite{Hamed1}-\cite{Hamed2} with possible combinations among them. Among the most economical ones is the top condensate model \cite{Cohen}-\cite{Chivukula} where it is envisioned that the top quark produces a  strong enough condensate that leads to the known electroweak vacuum through a version or another of the standard technicolor mechanisms.

In this work we will use an odd looking scalar operator extracted from the fermions equation of motion and the known fact that at least in first order the standard model is the correct theory of the electroweak interaction to extract the vacuum condensates of all quarks. In this process we do not refer at all to any strong processes but mostly use precise calculations to determine how and if these condensates contribute to the electroweak vacuum. We find out that standard mechanisms of the Nambu Jona-Lasinio type cannot explain the breaking of the electroweak scale. If the Higgs is to be a bound function of the quark states  then in order to account for the electroweak breaking it should have a form similar to that of the operator  described in the present work  showing that a possible  beyond standard model theory  might be completely new and unexpected by our standards.

Section II introduces the fermion operator of interest and its role in the electroweak vacuum. Section III contains detailed calculations of the vacuum value of this operator in the path integral formalism and its relation to the quarks vacuum condensates.  In section IV we discuss the numerical results and enounce our conclusions.

\section{The standard model Higgs boson}

We consider the standard model at a scale slightly below the electroweak symmetry breaking scale where all fermions and gauge bosons have masses. We are mainly interested in the fermion part of the Lagrangian, namely that associated with the Higgs boson:
\begin{eqnarray}
{\cal L}_f=\bar{\Psi}_f(i\gamma^{\nu}\partial_{\mu}+y_fB(x))\Psi_f+......
\label{fermlagr}
\end{eqnarray}
where $B$ is the neutral Higgs boson such that $B=\frac{1}{\sqrt{2}}(v+H)$  and all other interaction terms are neglected.  Here $f$ is the fermion flavor excluding the neutrinos and we assume without any loss of generality that the Yukawa constants are real and there is no important mixings between the fermions mass eigensstates.

We apply the equation of motion for the fermion fields to get:
\begin{eqnarray}
i\gamma^{\mu}\partial_{\mu}\Psi_f+y_fB\Psi_f+...=0
\label{motion36475}
\end{eqnarray}
or by multiplying with $\bar{\Psi}_f$,
\begin{eqnarray}
\bar{\Psi}_fi\gamma^{\mu}\partial_{\mu}\Psi_f+y_fB\bar{\Psi}_f\Psi_f+...=0.
\label{compl7484657}
\end{eqnarray}
Eq. (\ref{motion36475}) is in general employed for extracting fermion solutions of the equation of motion. Let us instead use Eq. (\ref{compl7484657}) for determining $B$. Thus we shall consider the following  change of variable of the
Higgs boson:
\begin{eqnarray}
B(x)=-\epsilon\frac{i\bar{\Psi}_f(x)\gamma^{\mu}\partial_{\mu}\Psi_f(x)}{y_f\bar{\Psi}_f(x)\Psi_f(x)}+\frac{1}{\sqrt{2}}h
\label{bchnage36}
\end{eqnarray}
Here $\Psi_f$ is a generic quark of  flavor $f$ (the summation over the number of colors in the numerator and denominator is understood) and $\epsilon$ is a positive parameter of arbitrary value. We will show later that the unusual operator in the first term of Eq. (\ref{bchnage36}) makes perfect sense in a quantum field theory. There are many possible scenarios of the same type from which we consider only another one which corresponds to the change of variable:
\begin{eqnarray}
B(x)=-\epsilon\sum_f\frac{i\bar{\Psi}_f\gamma^{\mu}\partial_{\mu}\Psi_f}{y_f\bar{\psi}_f\Psi_f}+\frac{1}{\sqrt{2}}h
\label{newch6483765}
\end{eqnarray}

We first introduce the change of variable in Eq. (\ref{bchnage36}) into Eq. (\ref{fermlagr})  to obtain:
\begin{eqnarray}
&&{\cal L}=\sum_g\bar{\Psi}_g(i\gamma^{\mu}\partial_{\mu})\Psi_g+\sum_g(-\epsilon\frac{i\bar{\Psi}_f\gamma^{\mu}\partial_{\mu}\Psi_f}{y_f\bar{\Psi}_f\Psi_f}+\frac{1}{\sqrt{2}}h)y_g\bar{\Psi}_g\Psi_g=
\nonumber\\
&&(1-\epsilon)\bar{\Psi}_f(i\gamma^{\mu}\partial_{\mu})\Psi_f+\sum_{g\neq f}\bar{\Psi}_g(i\gamma^{\mu}\partial_{\mu})\Psi_g-\epsilon\frac{i\bar{\Psi}_f\gamma^{\mu}\partial_{\mu}\Psi_f}{y_f\bar{\Psi}_f\Psi_f}\sum_{g\neq f}y_g\bar{\Psi}_g\Psi_g+\frac{1}{\sqrt{2}}hy_g\sum_g\bar{\Psi}_g\Psi_g.
\label{res638465789}
\end{eqnarray}
Since the fermion $f$ kinetic term gained a scale factor we scale it back to the correct value by making the change of variable $\Psi_f(1-\epsilon)^{\frac{1}{2}}\rightarrow\Psi_f$ and $\bar{\Psi}_f(1-\epsilon)^{\frac{1}{2}}\rightarrow \bar{\Psi}_f$. Then Eq. (\ref{res638465789}) will become:
\begin{eqnarray}
&&{\cal L}=\sum_g\bar{\Psi}_gi\gamma^{\mu}\partial_{\mu}\Psi_g-\epsilon\frac{i\bar{\Psi}_f\gamma^{\mu}\partial_{\mu}\Psi_f}{y_f\bar{\Psi}_f\Psi_f}\sum_{g\neq f}y_g\bar{\Psi}_g\Psi_g+
\nonumber\\
&&\frac{1}{\sqrt{2}}y_fh\frac{1}{1-\epsilon}\bar{\Psi}_f\Psi_f+\frac{1}{\sqrt{2}}\sum_{g\neq f}hy_g\bar{\Psi}_g\Psi_g.
\label{res638465789}
\end{eqnarray}

The above lagrangian should lead to the correct free particles solution for the fermion fields and in consequence:
\begin{eqnarray}
&&\langle h\rangle= v(1-\epsilon)
\nonumber\\
&&\langle -\epsilon\frac{i\bar{\Psi}_f\gamma^{\mu}\partial_{\mu}\Psi_f}{y_f\bar{\Psi}_f\Psi_f}\rangle+\frac{1}{\sqrt{2}}v(1-\epsilon)=\frac{1}{\sqrt{2}}v
\nonumber\\
&&\langle -\frac{i\bar{\Psi}_f\gamma^{\mu}\partial_{\mu}\Psi_f}{\bar{y_f\Psi}_f\Psi_f}\rangle=\frac{1}{\sqrt{2}}v.
\label{result58497586}
\end{eqnarray}

Next we introduce the change of variable in Eq. (\ref{newch6483765}) into Eq. (\ref{fermlagr}) to obtain in the  same manner that:
\begin{eqnarray}
&&\langle-\epsilon\frac{1}{1-\epsilon}\sum_{g\neq  f}\frac{i\bar{\Psi}_g\gamma^{\mu}\partial_{\mu}\Psi_g}{y_g\bar{\Psi}_g\Psi_g}\rangle+\frac{1}{1-\epsilon}\langle \frac{1}{\sqrt{2}}h\rangle=\frac{1}{\sqrt{2}}v
\nonumber\\
&&\langle-\epsilon\sum_f\frac{i\bar{\Psi}_f\gamma^{\mu}\partial_{\mu}\Psi_f}{y_f\bar{\Psi}_f\Psi_f}\rangle+\langle \frac{1}{\sqrt{2}}h\rangle=\frac{1}{\sqrt{2}}v,
\label{firstsyste547389}
\end{eqnarray}
to get:
\begin{eqnarray}
\frac{1}{\sqrt{2}}\frac{1}{1-\epsilon}v-\frac{1}{1-\epsilon}\langle \frac{1}{\sqrt{2}}h\rangle-\langle-\epsilon\frac{1}{1-\epsilon}\frac{i\bar{\Psi}_f\gamma^{\mu}\partial_{\mu}\Psi_f}{y_f\bar{\Psi}_f\Psi_f}\rangle+\frac{1}{1-\epsilon}\langle \frac{1}{\sqrt{2}}h\rangle=\frac{1}{\sqrt{2}}v
\label{res774585}
\end{eqnarray}
which further leads to:
\begin{eqnarray}
\langle -\frac{i\bar{\Psi}_f\gamma^{\mu}\partial_{\mu}\Psi_f}{y_f\bar{\Psi}_f\Psi_f}\rangle=\frac{1}{\sqrt{2}}v.
\label{res637489}
\end{eqnarray}
The result in Eq. (\ref{res637489}) is thus quite generic.

Next we will show how one can deal with operator like that in Eq. (\ref{bchnage36}) in a quantum field theory. For that we consider a solution of the equation of motion of the free field of the type:
\begin{eqnarray}
\chi=\chi_++\chi_-,
\label{solfree53678}
\end{eqnarray}
where,
\begin{eqnarray}
\chi_+=\sqrt{m}\Lambda\exp[-imx_0]\times
\left(
\begin{array}{c}
1\\
0\\
1\\
0
\end{array}
\right)
\nonumber\\
\chi_=\sqrt{m}\Lambda\exp[-imx_0]\times
\left(
\begin{array}{c}
0\\
1\\
0\\
1
\end{array}
\right),
\label{expr75649230039}
\end{eqnarray}
where $\Lambda$ is a parameter with mass dimension 1 that we can take as large as we want.
Then we expand the fermion operator around this solution to get:
\begin{eqnarray}
&&-\frac{\Psi\gamma^{\mu}\partial_{\mu}\Psi}{y\bar{\Psi}\Psi}=
-\frac{\chi\gamma^{\mu}\partial_{\mu}\chi}{y\bar{\chi}\chi}\Bigg[1-\frac{\bar{\Psi}\Psi+\bar{\chi}\Psi+\bar{\Psi}\chi}{\bar{\chi}\chi}+....\Bigg]-
\nonumber\\
&&\frac{\bar{\Psi}i\gamma^{\mu}\partial_{\mu}\chi+\bar{\chi}i\gamma^{\mu}\partial_{\mu}\Psi+\bar{\Psi}\gamma^{\mu}\partial_{\mu}\Psi}{y\bar{\chi}\chi}\times
\Bigg[1-\frac{\bar{\Psi}\Psi+\bar{\chi}\Psi+\bar{\Psi}\chi}{\bar{\chi}\chi}+....\Bigg]=
\nonumber\\
&&\frac{1}{\sqrt{2}}v\Bigg[1-\frac{\bar{\Psi}\Psi+\bar{\chi}\Psi+\bar{\Psi}\chi}{4m\Lambda^2}+....\Bigg]-
\nonumber\\
&&\frac{\bar{\Psi}i\gamma^{\mu}\partial_{\mu}\chi+\bar{\chi}i\gamma^{\mu}\partial_{\mu}\Psi+\bar{\Psi}\gamma^{\mu}\partial_{\mu}\Psi}{y4m\Lambda^2}\times
\Bigg[1-\frac{\bar{\Psi}\Psi+\bar{\chi}\Psi+\bar{\Psi}\chi}{4m\Lambda^2}+....\Bigg].
\label{resu65748}
\end{eqnarray}
Here we used the equation of motion for the field $\chi$ and the fact that $\bar{\chi}\chi=4m\Lambda^2$. It is evident that we obtain an expansion in $\frac{1}{\Lambda}$ and since $\Lambda$ can be made arbitrarily large the higher dimension operators are very suppressed and can be ignored. Thus we showed that not only the odd looking operator makes sense but also that in first order its vacuum expectation value replaces that of the Higgs.

\section{Fermion vacuum condensates }

 We will estimate the fermion operator in Eq. (\ref{bchnage36}) in the path integral formalism in first order in the expansion of the Lagrangian. We will assume that the fermions (we discuss the quarks but the same arguments apply to the charged leptons) have a scaled vacuum condensate $z_f'$ (see  Appendix A). Then one can write:
\begin{eqnarray}
&& T=-\frac{\bar{\Psi}i\gamma^{\mu}\partial_{\mu}\Psi}{y\bar{\Psi}\Psi}=
 -\frac{1}{y}\frac{\bar{\Psi}i\gamma^{\mu}\partial_{\mu}\Psi}{z'+\bar{\Psi\Psi}}=
\nonumber\\
&&-\frac{1}{yz'}\bar{\Psi}i\gamma^{\mu}\partial_{\mu}\Psi\Bigg[1-\frac{\bar{\Psi}\Psi}{z'}+\frac{(\bar{\Psi}\Psi)^2}{z^{\prime 2}}-...\Bigg],
\label{resul74637456}
\end{eqnarray}
where we dropped the index $f$ corresponding to the fermion flavor for the simplicity of the equations.

Then in the path integral formalism the vacuum expectation value of the fermion operator is given by:
\begin{eqnarray}
\langle T \rangle=
\frac{1}{Z}\int d\bar{\Psi}d \Psi T(x)\exp[i\int d^4 x{\cal L}].
\label{expr743836}
\end{eqnarray}
Here in ${\cal L}$ we include only the fermion kinetic term because we are interested in a first order estimate and $Z$ is the partition function. We first denote:
\begin{eqnarray}
&&T=-\frac{U(x)}{yz'}\Bigg[1-\frac{V(x)}{z'}+\frac{V(x)^2}{z^{\prime 2}}-...\Bigg]
\nonumber\\
&&Y_n(x)=-\frac{U(x)}{yz^{\prime n+1}}V(x)^n(-1)^n
\label{fomr75847657}
\end{eqnarray}
where,
\begin{eqnarray}
&&U(x)=\bar{\Psi}(x)i\gamma^{\mu}\partial_{\mu}\Psi(x)
\nonumber\\
&&V(x)=\bar{\Psi}(x)\Psi(x).
\label{res6295748}
\end{eqnarray}

We shall estimate the operators in Eqs.(\ref{expr743836}), (\ref{fomr75847657}) in the path integral formalism and rely heavily on the mathematical results regarding fermion integration obtained in \cite{Feldman}.
These results are exact in the limit where $m\ll\Lambda$ where $\Lambda$ is the cut-off of the theory.
First we will use the fact that integration by parts apply as well to the integration of the Grassmann variables. This leads to:
\begin{eqnarray}
&&\frac{1}{Z}\int d\bar{\Psi} d\Psi (\bar{\Psi}(x)\Psi(x))^n\exp[i\int d^4 x {\cal L}]=
\nonumber\\
&&\int d \bar{\Psi}d\Psi\frac{1}{12\delta(0)}\sum_k\frac{\delta \bar{\Psi}_k(x)}{\delta \bar{\Psi}_k(x)}(\bar{\Psi}(x)\Psi(x))^n\exp[i\int d^4 x{\cal L}]=
\nonumber\\
&&-\int  d\bar{\Psi}d\Psi \frac{n}{\delta(0)12}\delta(0)(\bar{\Psi}(x)\Psi(x))^n\exp[i\int d^4 x{\cal L}]-
\nonumber\\
&&-\int d\bar{\Psi} d\Psi \frac{1}{\delta(0)12}(\bar{\Psi}(x)\Psi(x))^n\bar{\Psi}(x)i\gamma^{\mu}\partial_{\mu}\Psi(x)\exp[i\int d^4 x{\cal L}].
\label{res73648995}
\end{eqnarray}
Here $12$ corresponds to the sum over space time and color degrees of freedom
Eq. (\ref{res73648995})  yields to the following recurrence relation:
\begin{eqnarray}
&&\langle (\bar{\Psi}(x)\Psi(x))^n\rangle=-\frac{n}{12}\langle (\bar{\Psi}(x)\Psi(x))^n\rangle+\frac{1}{12\delta(0)}(-1)^nyz^{\prime n+1}\langle Y_n(x)\rangle
\nonumber\\
&&\langle Y^n\rangle=12\delta(0)(-1)^n [1+\frac{n}{12}]\frac{1}{yz^{\prime n+1}}\langle V^n(x)\rangle
\label{res663547}
\end{eqnarray}

We shall consider the last equation in Eq. (\ref{res663547}) as the fundamental relation applicable for $n\geq1$.

The next step is to determine the vacuum correlator:
\begin{eqnarray}
&& \frac{1}{Z}\int \bar{d\Psi}d\Psi(-1)^n(\bar{\Psi}(x)\Psi(x))^n\exp[i\int d^4 x {\cal L}]=
\nonumber\\
&&(-1)^n(-1)^n\int\prod_{k=1}^n\frac{d^4p_k}{(2\pi)^4}\sum_{i_1..i_n}\times
\nonumber\\
&&\det
\left[
\begin{array}{ccc}
(\frac{1}{\gamma^{\mu}p_{1\mu}-m})_{i_1i_1}&...&(\frac{1}{\gamma^{\mu}p_{1\mu}-m})_{i_1i_n}
\\
...&...&...
\\
(\frac{1}{\gamma^{\mu}p_{n\mu}-m})_{i_ni_1}&...&(\frac{1}{\gamma^{\mu}p_{n\mu}-m})_{i_ni_n}
\end{array}
\right]=
 \nonumber\\
 &&(-1)^n(-1)^n\sum_{i_1...i_n}\det[\delta_{i_k,i_j}]\times\Bigg[\frac{m}{16\pi^2}(\Lambda^2-m^2Log[\frac{\Lambda^2+m^2}{m^2}])\Bigg]^n
  \label{res84637}
 \end{eqnarray}
 Note that the indices $i_i$...$i_n$ go over the four space time degrees of freedom and three degrees of color and also that in second line we performed the integration over the momentum variables using the standard procedure with  a cut-off $\Lambda$. The sum of the determinants can be calculated easily if one considers the contractions in the path integral formalism. We will prove this by induction. Thus it is evident that for $n=1$ , $\sum_{i_1...i_n}\det[\delta_{i_k,i_j}]=12$ and for $n=2$ the same sum is $12\times 11$ (this can be seen by simple calculations). Then we assume that for $n$ we have:
 \begin{eqnarray}
\sum_{i_1,...i_n}\det[\delta_{ik,ij}]=(3\times4)(3\times4-1)....(3\times 4-(n-1))
\label{ordreind6574895}
\end{eqnarray}
We need to show:
 \begin{eqnarray}
\sum_{i_1,...i_{n+1}}\det[\delta_{i_k,i_j}]=(3\times4)(3\times4-1)....(3\times 4-(n))
\label{ordreind6574895}
\end{eqnarray}
We shall prove this in terms of contractions in the group of fermions $(\bar{\Psi}\Psi)^{n+1}$ as we separate the group $\bar{\Psi}\Psi$ and the group $(\bar{\Psi}\Psi)^n$. First we consider contraction within each group which leads to $-(-1)^n12(12-1)...(12-n-1)$. Then we consider that one $\bar{\Psi}$ within the first group contracts with one $\Psi$ from the second group. Then there are n possible contractions and we know the number of contraction for the rest of the group so we get $n(-1)^n(12)...(12-n-1)$. The final result is thus $(-1)^{n+1}12...(12-n)$.

Also this can be regarded as  the correspnding integral over arbitrary Grassmann variables $\eta_k$ and $\bar{\eta_k}$:
\begin{eqnarray}
&&\int d\bar{\eta}_k\eta_k\exp[-\bar{\eta}_j\eta_j a]=a^m
\nonumber\\
&&\frac{1}{a^m}\int d\bar{\eta}_k\eta_k (\bar{\eta}\eta)^n\exp[\bar{\eta}_j\eta_j a]=
\nonumber\\
&&(-1)^n\frac{1}{a^m}\frac{\delta}{\delta^n a}a^m=(-1)^n\sum_{i_r} \det[\delta_{i_{r=1...n},i_{s=1...n}}]a^{n-m}=(-1)^nm(m-1)...(m-n+1)a^{n-m}
\label{imort657489}
\end{eqnarray}

First we notice that the sum of interest stops at $n=12$ so it can be easily computed numerically.
Then we infer from Eqs. (\ref{res84637}) and (\ref{fomr75847657}):
\begin{eqnarray}
&&T=-\frac{1}{yz'}\langle \bar{\Psi}i\gamma^{\mu}\partial_{\mu}\Psi\rangle+\sum_{n=1}^{12}12\delta(0)(-1)^n [1+\frac{n}{12}]\frac{1}{yz^{\prime n+1}}\langle V^n(x)\rangle=
\nonumber\\
&&-\frac{1}{yz'}\frac{3}{8\pi^2}[\Lambda^4-2m^2\Lambda^2+2m^4Log[\frac{\Lambda^2+m^2}{m^2}]]+\sum_{n=1}^{12}\frac{3}{8\pi^2}\Lambda^4[1+\frac{n}{12}]\times
\nonumber\\
&&\frac{1}{yz^{\prime n+1}}(12(12-1)....(12-n+1))[\frac{m}{16\pi^2}(\Lambda^2-m^2Log[\frac{\Lambda^2+m^2}{m^2}])]^n
\label{finalres65747}
\end{eqnarray}

Knowing that in first order $\langle T\rangle=\frac{v}{\sqrt{2}}$ we use $y\frac{v}{\sqrt{2}}=-m$ to obtain the final equality:
\begin{eqnarray}
&&1=\frac{1}{mz'}\frac{3}{8\pi^2}[\Lambda^4-2m^2\Lambda^2+2m^4Log[\frac{\Lambda^2+m^2}{m^2}]]-\frac{1}{mz'}\frac{3}{8\pi^2}\Lambda^4\times
\nonumber\\
&&\sum_{n=1}^{12}[1+\frac{n}{12}](12(12-1)....(12-n+1))[\frac{m}{16\pi^2z'}(\Lambda^2-m^2Log[\frac{\Lambda^2+m^2}{m^2}])]^n.
\label{final7493999}
\end{eqnarray}
Note that in the second term for heavy quarks the factor $\Lambda^4$ should be replaced by the quantity $[\Lambda^4-2m^2\Lambda^2+2m^4Log[\frac{\Lambda^2+m^2}{m^2}]]$ to take into account the effect of the masses.

\section{Numerical results and conclusions}

In this section we will apply Eq.(\ref{final7493999}) for different scales to determine the magnitude of the quark condensates. First we notice that the operator in Eq. (\ref{resul74637456}) must be introduced in the Lagrangian through the change of variables in Eq. (\ref{bchnage36}). Then the fermion field get scaled by $\frac{1}{(1-\frac{\epsilon'}{yz})^{1/2}}$ where $\epsilon'=\epsilon\sum_g(y_gz_g)$ (see Appendix A) and  $z'$  is related to the true fermion condensate $z$ by the relation $z'=(1-\frac{\epsilon'}{yz})z$. Since this change of variables introduces an additional parameter $\epsilon'$  this will get renormalized. We will identify the cut-off scale with the renormalization one such that the masses and the other parameters that appear in Eq. (\ref{resul74637456}) correspond to the renormalized ones.

In Table \ref{lightquarks1} we display the  down and strange quark vacuum condensates as obtained in this work for $\Lambda=2$ GeV  and for two values of the parameter $\epsilon'$ extracted from two possible choices for the up quark condensates: $(z_{0u})_1=-0.016$ ${\rm GeV}^3$ \cite{Jora} and $(z_{0u})_2=-0.283^3$ ${\rm GeV}^3$ respectively $\epsilon_1'=2.08\times 10^{-7}$ and $\epsilon_2'=2.96\times 10^{-7}$.  We employ the following quark masses \cite{PDG}: $m_u=2.3\times 10^{_3}$ GeV, $m_d=4.8\times 10^{-3}$ GeV and $m_s=95\times 10^{-3}$ GeV.
\begin{table}[htbp]
\begin{center}
\begin{tabular}{c|c|c}
\hline \hline  &${\rm Down\, quark\, condensate \,\,\,}GeV^3 $&${\rm Strange\, quark\, condensate \,\,\,}GeV^3$
\\
\hline \hline $\epsilon_1'=2.08\times 10^{-7}$&$-0.008$&$-0.001$
\\
\hline\hline $\epsilon_2'=2.96\times 10^{-7}$&$-0.011$&$-0.001$
\\
\hline\hline
\end{tabular}
\end{center}
\caption[Theoretical estimate]{Vacuum condensates for the down and strange quarks for two choices of the vacuum condensate of the up quark: $0.016\,\,\,GeV^3$ ($\epsilon_1=2.08\times 10^{-7}$) and $0.283^3\,\,\,GeV^3$ ($\epsilon_2=2.08\times 10^{-7}$). Here $\Lambda=2$ GeV. }
\label{lightquarks1}
\end{table}

 In Table \ref{heavyquarks} we present the heavy quark cotribution to the vacuum as determined for $\Lambda=246.22$ GeV for the two values for $\epsilon'$ determined above, $\epsilon_1'$ and $\epsilon_2'$.   We use the known values of heavy quark masses; $m_c=1.275 $ GeV, $m_b=4.18$ GeV and $m_t=173.21$ GeV. We make the assumption that the parameter $\epsilon'$ runs slowly with the scale.

\begin{table}[htbp]
\begin{center}
\begin{tabular}{c|c|c|c|c|c}
\hline \hline  &${\rm Charm\, quark\, condensate \,\,\,}GeV^3$ &${\rm Bottom\, quark\, condensate \ \,\,\,}GeV^3 $&${\rm Top\, quark\, condensate \  \,\,\,}GeV^3$
\\
\hline \hline $\epsilon_1'=2.08\times 10^{-7}$&$2014.54$&$6590.87$&$116596.00$
\\
\hline\hline $\epsilon_2'=2.96\times 10^{-7}$&$2014.54$&$6590.87$&$116596.00$
\\
\hline\hline
\end{tabular}
\end{center}
\caption[Theoretical estimate]{Heavy quark condensates computed at  $\Lambda -246.22$ GeV for the two values $\epsilon_1$ and $\epsilon_2$. }
\label{heavyquarks}
\end{table}

The light quarks  vacuum condensates decrease with the masses whereas the heavy quark condensates increase with the masses.
It may appear for example that such a large condensate for the top quark (as that for the scale $\Lambda=246.22$ GeV)  might lead to a significant contribution of the top quark condensate to the electroweak vacuum. However this is not true. The real contribution to the electroweak breaking is given by $\frac{z_t}{\Lambda^2}=\frac{\bar{\Psi}_t\Psi_t}{\Lambda^2}$ which for example is of order $2$ GeV at the electroweak scale. This shows that the top quark condensate is not a decisive factor in  electroweak symmetry breaking at least not through a Nambu Jona-Lasinio type of mechanism.

In conclusion, based on  simple change of variable in the Higgs field we determine that the intricate fermion operator introduced in Eq. (\ref{resul74637456}) is entirely responsible for the Higgs vacuum expectation value. By computing this operator in the path integral formalism we established a relation among the quark vacuum condensates, the running masses and the renormalization scale. The vacuum condensates for the heavy quarks are relatively high but their associated scalar vacuum expectation values do not justify a picture in which the electroweak vacuum is broken by a simple technicolor type interaction. Our results support the idea that whereas a strong mechanism may be at play in the electroweak symmetry breaking, its implementation is far more complicated than the simple topcolor or technicolor picture. If the Higgs is to be composite of quark states we expect that its wave function is  more complex than that of a standard two fermion bound state. Our findings are based on the assumption, justified by the latest LHC results that the standard model as we know is at least in first order the best description of the electroweak scale and its particles and phenomena.

\begin{appendix}
\section{}

We consider the change of variable in Eq. (\ref{bchnage36}) but the same arguments apply as well to Eq. (\ref{newch6483765}).  We assume that all quark flavors have vacuum condensates denoted by $z_f$ where $f$ is the flavor and $z_f$ might have any value including zero. We introduce the change of variables into the Lagrangian for the fermion fields and take into account the vacuum condensates:

\begin{eqnarray}
&&{\cal L}=\sum_g\bar{\Psi}_g(i\gamma^{\mu}\partial_{\mu})\Psi_g+\sum_g(-\epsilon\frac{i\bar{\Psi}_f\gamma^{\mu}\partial_{\mu}\Psi_f}{y_f\bar{\Psi}_f\Psi_f}+\frac{1}{\sqrt{2}}h)y_g\bar{\Psi}_g\Psi_g=
\nonumber\\
&&(1-\epsilon)\bar{\Psi}_f(i\gamma^{\mu}\partial_{\mu})\Psi_f+\sum_{g\neq f}\bar{\Psi}_g(i\gamma^{\mu}\partial_{\mu})\Psi_g-\epsilon\frac{i\bar{\Psi}_f\gamma^{\mu}\partial_{\mu}\Psi_f}{y_f\bar{\Psi}_f\Psi_f}\sum_{g\neq f}y_g\bar{\Psi}_g\Psi_g+\frac{1}{\sqrt{2}}hy_g\sum_g\bar{\Psi}_g\Psi_g=
\nonumber\\
&&(1-\epsilon)\bar{\Psi}_f(i\gamma^{\mu}\partial_{\mu})\Psi_f+\sum_{g\neq f}\bar{\Psi}_g(i\gamma^{\mu}\partial_{\mu})\Psi_g-
\nonumber\\
&&\epsilon\frac{i\bar{\Psi}_f\gamma^{\mu}\partial_{\mu}\Psi_f}{y_f\bar{\Psi}_f\Psi_f}\sum_{g\neq f}(y_gz_g+y_g\bar{\Psi}_g\Psi_g)+\frac{1}{\sqrt{2}}h(y_gz_g+y_g\sum_g\bar{\Psi}_g\Psi_g)=
\nonumber\\
&&(1-\epsilon)\bar{\Psi}_f(i\gamma^{\mu}\partial_{\mu})\Psi_f+\sum_{g\neq f}\bar{\Psi}_g(i\gamma^{\mu}\partial_{\mu})\Psi_g-
\nonumber\\
&&\epsilon\frac{i\bar{\Psi}_f\gamma^{\mu}\partial_{\mu}\Psi_f}{y_f\bar{\Psi}_f\Psi_f}\sum_{g\neq f}(y_gz_g)-
\epsilon\frac{i\bar{\Psi}_f\gamma^{\mu}\partial_{\mu}\Psi_f}{y_f\bar{\Psi}_f\Psi_f}\sum_{g\neq f}(y_g\bar{\Psi}_g\Psi_g)+\frac{1}{\sqrt{2}}h(y_gz_g+y_g\sum_g\bar{\Psi}_g\Psi_g).
\label{res43526}
\end{eqnarray}

Then we calculate:
\begin{eqnarray}
-\epsilon\frac{i\bar{\Psi}_f\gamma^{\mu}\partial_{\mu}\Psi_f}{y_f\bar{\Psi}_f\Psi_f}\sum_{g\neq f}(y_gz_g)=
-\epsilon\frac{i\bar{\Psi}_f\gamma^{\mu}\partial_{\mu}\Psi_f}{y_fz_f}\sum_{g\neq f}(y_gz_g)+\frac{\bar{\Psi}_f\Psi_f}{z_f}\sum_{g\neq f}(y_gz_g)\epsilon\frac{i\bar{\Psi}_f\gamma^{\mu}\partial_{\mu}\Psi_f}{y_f\bar{\Psi}_f\Psi_f}.
\label{res63728}
\end{eqnarray}

From Eqs. (\ref{res43526}) and (\ref{res63728}) we extract the factor in front of the kinetic term for the flavor $f$ as: $1-\epsilon\frac{\sum_g y_gz_g}{y_fz_f}$. We perform the change of variable $\Psi_f\rightarrow 
\Psi_f\frac{1}{(1-\epsilon\frac{\sum_g y_gz_g}{y_fz_f})^{\frac{1}{2}}}$ and $\bar{\Psi}_f\rightarrow
\bar{\Psi}_f\frac{1}{(1-\epsilon\frac{\sum_g y_gz_g}{y_fz_f})^{\frac{1}{2}}}$ and  impose the equation of motion in first order for all fermion fields.  This leads to two conditions:
\begin{eqnarray}
&&\frac{\epsilon}{y_fz_f}\frac{1}{1-\epsilon\frac{\sum_g y_gz_g}{y_fz_f}}\sum_{g\neq f}(y_gz_g)
\langle\epsilon\frac{i\bar{\Psi}_f\gamma^{\mu}\partial_{\mu}\Psi_f}{y_f\bar{\Psi}_f\Psi_f}\rangle +\frac{1}{\sqrt{2}}\langle h\rangle\frac{1}{1-\epsilon\frac{\sum_g y_gz_g}{y_fz_f}}=\frac{v}{\sqrt{2}}
\nonumber\\
&&-\langle\epsilon\frac{i\bar{\Psi}_f\gamma^{\mu}\partial_{\mu}\Psi_f}{y_f\bar{\Psi}_f\Psi_f}\rangle+\frac{1}{\sqrt{2}}\langle h\rangle=\frac{v}{\sqrt{2}}.
\label{cond648375679}
\end{eqnarray}
and further yields:
\begin{eqnarray}
\langle -\frac{i\bar{\Psi}_f\gamma^{\mu}\partial_{\mu}\Psi_f}{y_f\bar{\Psi}_f\Psi_f}\rangle=\frac{v}{\sqrt{2}}.
\label{finalrel74857658}
\end{eqnarray}
We denote $\epsilon'=\epsilon\sum_g(y_gz_g)$ and by $z_f'=(1-\frac{\epsilon'}{y_fz_f})z_f$ the scaled vacuum condensate that appears in the denominator of the operator in Eq. (\ref{finalrel74857658}). Then the 
dimensionless parameter $\epsilon'$ will get renormalized.

\end{appendix}

\section*{Acknowledgments} \vskip -.5cm

The work of R. J. was supported by a grant of the Ministry of National Education, CNCS-UEFISCDI, project number PN-II-ID-PCE-2012-4-0078.


\begin{thebibliography}{15}
\bibitem{Atlas} G. Aad {\it et al.}[ATLAS Collaboration], Phys. Lett. B {\bf 716}, 1 (2012)
\bibitem{CMS} S. Chatrchyan {\it et al.} [CMS Collaboration], Phys. Lett. B {\bf 716}, 30 (2012).
\bibitem{Weinberg1} S. Weinberg, Phys. Rev.  {\bf D  13}, 974 (1976).
\bibitem{Weinberg2} S. Weinberg, Phys. Rev.  {\bf D 19}, 1277 (1979).
\bibitem{Susskind} L. Susskind, Phys. Rev.  {\bf  D 20}, 2619 (1979).
\bibitem{Kaplan1} D. B. Kaplan and H. Georgi, Phys. Lett.  {\bf B 136}, 183 (1984).
\bibitem{Kaplan2} D. B. Kaplan, H. Georgi and S. Dimopoulos, Phys. Lett.  {\bf B 136}, 187 (1984).
\bibitem{Peskin} M. E. Peskin, Nucl. Phys.  {\bf B 175}, 197 (1980).
\bibitem{Preskill} J. Preskill, Nucl. Phys.  {\bf B 177}, 21 (1981).
\bibitem{Cohen} R. S Chivukula, A. G. Cohen and K. D. Lane, Nucl. Phys.  {\bf B 343}, 554 (1990).
\bibitem{Csaki} B. Bellazzini, C. Csaki and J. Serra, Eur. Phys. J.  {\bf C 74}, 2766 (2014).
\bibitem{Miransky} V. A. Miransky, M. Tanabashi and K. Yamawaki, Phys. Lett.  {\bf B 221}, 177 (1989); Mod. Phys. Lett.  {\bf A 4}, 1043 (1989).
\bibitem{Bardeen} W. A. Bardeen, C. T. Hill and M. Lindner, Phys. Rev.  {\bf D 41}, 1647 (1990).
\bibitem{Hill} C. T. Hill, Phys. Lett.  {\bf B 266}, 419 (1991).
\bibitem{Dobrescu} B. A. Dobrescu and C. T. Hill, Phys. Rev. Lett {\bf 81}, 2634 (1998).
\bibitem{Chivukula} R. Chivukula et al., Phys. Rev.  {\bf D 59}, 075003 (1999).
\bibitem{Nilles}H. P. Nilles, Phys. Reports {\bf 110}, 1 (1984).
\bibitem{Witten} E. Witten, NUcl. Phys.  {\bf B 188}, 513 (1981).
\bibitem{Georgi} S. Dimopoulos and H. Georgi, Nucl. Phys.  {\bf B 193}, 150 (1981).
\bibitem{Sakai} N. Sakai, Z. Phys.  {\bf C 11}, 153 (1981).
\bibitem{Kaul1} R. K. Kaul, Phys. Lett. {\bf 109 B }, 19 (1982).
\bibitem{Kaul2} R. K. Kaul and M. Parthasarathi, Nucl. Phys.  {\bf B 199}, 36 (1982).
\bibitem{Susskind2} L. Susskind. Phys. Reports {\bf 104}, 181 (1984).
\bibitem{Grisaru} L. Girardello and M. Grisaru, Nucl. Phys.  {\bf B 194}, 65 (1982).
\bibitem{Hall} L. J. Hall and L. Randall, Phys. Rev. Lett. {\bf 65}, 2939 (1990).
\bibitem{Jones} I. Kack and D. R. T. Jones, Phys. Lett.  {\bf B 457}, 101 (1999).
\bibitem{Kane} H. E. Haber and G. L. Kane, Phys. Reoprts {\bf 117}, 75 (1985).
\bibitem{Hamed1} N. Arkani-Hamed et al., Phys. Lett {\bf B 429}, 263 (1998).
\bibitem{Randall} L. Randall and R. Sundrum, Phys. Rev. Lett. {\bf 83}, 3370 (1999).
\bibitem{Davoudiasl} H. Davoudiasl et al,,Phys. Lett. {\bf B 473}, 43 (2000).
\bibitem{Pomarol} A. Pomarol, Phys. Lett. {\bf B 486}, 153 (2000).
\bibitem{Chang} S. Chang et al., Phys. Rev. {\bf D 62}, 084025 (2000).
\bibitem{Pomarol1} T. Gherghetta and A. Pomarol, Nucl. Phys. {\bf B 586}, 141 (2000).
\bibitem{Hamed2} N. Arkani-Hamed et al., JHEP {\bf 0108}, 017 (2001).
\bibitem{Feldman} J. Feldman, H. Kn${ \rm\ddot{o}}$rrer and E. Trubowitz, "Fermionic Functional Integrals and the Renormalization Group", CRM Monograph Series 16, Providence RI; American Mathematical Society.
\bibitem{Jora} A. H. Fariborz, R. Jora and J. Schechter, Phys. Rev. {\bf D 77}, 094004 (2008).
\bibitem{PDG} C. Patrignani et al. (Particle  Data Group), Chin. Phys. C {\bf 40}, 100001 (2016).


\end{thebibliography}
\end{document}